\documentclass[%
 reprint,
superscriptaddress,
 amsmath,amssymb,
 aps,
]{revtex4-2}

\usepackage{graphicx}% Include figure files
\usepackage{dcolumn}% Align table columns on decimal point
\usepackage{bm}% bold math
\usepackage{xcolor}

\begin{document}

\preprint{APS/123-QED}

\title{Complementary Vanadium Dioxide Metamaterial with Enhanced  Modulation Amplitude at Terahertz Frequencies}% Force line breaks with \\

\author{Yuwei Huang}

\altaffiliation{These two authors contributed equally}
\affiliation{Department of Mechanical Engineering and Division of Materials Science and Engineering, Boston University, Boston, Massachusetts 02215, USA}
\author{Xuefei Wu}
\altaffiliation{These two authors contributed equally}
\affiliation{Department of Mechanical Engineering and Division of Materials Science and Engineering, Boston University, Boston, Massachusetts 02215, USA}
\author{Jacob Schalch}
\affiliation{Department of Physics, University of California, San Diego, La Jolla, California 92903, USA}
\author{Guangwu Duan}
\affiliation{Department of Mechanical Engineering and Division of Materials Science and Engineering, Boston University, Boston, Massachusetts 02215, USA}
\author{Chunxu Chen}
\affiliation{Department of Mechanical Engineering and Division of Materials Science and Engineering, Boston University, Boston, Massachusetts 02215, USA}
\author{Xiaoguang Zhao}
\affiliation{Department of Mechanical Engineering and Division of Materials Science and Engineering, Boston University, Boston, Massachusetts 02215, USA}
\author{Kelson Kaj}
\affiliation{Department of Physics, University of California, San Diego, La Jolla, California 92903, USA}
\author{Hai-Tian Zhang}
\affiliation{Department of Materials Science and Engineering, the Pennsylvania State University, University Park, Pennsylvania 16802, USA}
\author{Roman Engel-Herbert}
\affiliation{Department of Materials Science and Engineering, the Pennsylvania State University, University Park, Pennsylvania 16802, USA}
\affiliation{Paul-Drude-Institut für Festkörperelektronik, Leibniz-Institut im Forschungsverbund Berlin e.V., Hausvogteiplatz 5-7, 10117 Berlin, Germany}
\author{Richard D. Averitt}
\email{raveritt@ucsd.edu}
\affiliation{Department of Physics, University of California, San Diego, La Jolla, California 92903, USA}
\author{Xin Zhang}
\email{xinz@bu.edu}
\affiliation{Department of Mechanical Engineering and Division of Materials Science and Engineering, Boston University, Boston, Massachusetts 02215, USA}

\date{\today}% It is always \today, today,

\begin{abstract}
One route to create tunable metamaterials is through integration with “on-demand” dynamic quantum materials, such as vanadium dioxide (VO$_{2}$). This enables new modalities to create high performance devices for historically challenging applications. Indeed, dynamic materials have often been integrated with metamaterials to imbue artificial structures with some degree of tunability. Conversely, metamaterials can be used to enhance and extend the natural tuning range of dynamic materials. Utilizing a complementary split ring resonator array deposited on a VO$_{2}$ film, we demonstrate enhanced terahertz transmission modulation upon traversing the insulator-to-metal transition (IMT) at $\sim$340 K. Our complementary metamaterial increases the modulation amplitude of the original VO$_{2}$ film from 42\% to 68.3\% at 0.47 THz upon crossing the IMT, corresponding to  an enhancement of 62.4\%. Moreover, temperature dependent transmission measurements reveal a redshift of the resonant frequency arising from a giant increase of the 
permittivity of the VO$_{2}$ film. Maxwell-Garnett effective medium theory was employed to explain the permittivity change upon transitioning through the IMT. Our results highlight that symbiotic integration of metamaterial arrays with quantum materials provides a powerful approach to engineer emergent functionality.
%\begin{description}
%\end{description}
\end{abstract}

\maketitle

%\tableofcontents

\section{\label{sec:level1}Introduction}
Metamaterials, and their single layer analogs (metasurfaces), have enabled a plethora of applications and devices by overcoming the limitations of natural materials. This includes subwavelength imaging \cite{watts2014terahertz}, the realization of a negative refractive index \cite{smith2004metamaterials}, near zero epsilon devices \cite{maas2013experimental}, and the demonstration of cloaks and metalenses \cite{ni2015ultrathin,khorasaninejad2017metalenses}. Building on these capabilities, active control and tunability in metamaterials could lead to novel modulators, switches, and sensors over a vast span of the electromagnetic spectrum \cite{chen2006active,zhao2018electromechanically,liu2012terahertz,fan2013nonlinear,huang2022broadband}.

Vanadium dioxide (VO$_{2}$) is a tunable quantum material that has been extensively investigated (in large part) because the insulator-to-metal transition (IMT) occurs above room temperature at $\sim$340 K \cite{morin1959oxides}. Light, current, and mechanical strain can also drive the IMT, making VO$_{2}$ useful for a broad range of applications \cite{liu2012terahertz,cao2009strain,o2015inhomogeneity,jeong2013suppression,tao2012decoupling,zhu2022vo2}. Upon undergoing the IMT, the optical conductivity changes by four-orders of magnitude over a broad spectral range, making VO$_{2}$ a versatile material to integrate with metamaterials to realize nonlinear responses, state switches, and modulators \cite{liu2016hybrid,cueff2015dynamic}. Recent efforts have focused on combining VO$_{2}$ with metamaterials in the terahertz (THz) frequency range to realize switchable broadband absorbers, switchable coding and multi-functional devices based on the IMT \cite{li2020frequency,li2022switchable,shabanpour2020ultrafast,ding2018vanadium}.

In order to enhance the THz response and explore the phase transition behavior of VO$_{2}$, we deposited a metamaterial array of complementary split ring resonators (SRRs) on a thin film using direct laser writing and photolithography. As described below, the resultant structure exhibits a significantly enhanced THz transmission modulation across the IMT in excess of a pristine VO$_{2}$ film. Previous studies of transmission modulation suffer from a high insertion loss ($>$ 40\%) \cite{seo2010active} or a relatively small amplitude modulation enhancement in comparison to a pristine film \cite{zhu2022vo2}. 

In addition to the functional increase of the THz transmission modulation upon traversing the IMT, the sensitivity of the metamaterial resonators to the surrounding dielectric environment reveals an abnormally large increase of the permittivity of VO$_{2}$ during the IMT. This phenomenon can be attributed to insulating and metallic phase coexistence at the onset of the IMT, which has previously been studied by scanning near-field infrared microscopy \cite{qazilbash2007mott,qazilbash2009infrared,he2009high}. With increasing temperature, the formation of rutile metallic puddles surrounded by an insulating VO$_{2}$ matrix results in an increase in the effective permittivity, as described by Maxwell-Garnett effective medium theory (MG-EMT) \cite{garnett1906vii}. Although Bruggemen EMT has also been employed to explain the IMT \cite{liu2019vanadium}, our results indicate that MG-EMT better describes the dielectric properties of our integrated metamaterial VO$_{2}$ film, especially in the vicinity of the transition \cite{jepsen2006metal}.  

In the following, we describe the transmission enhancement properties of integrated metamaterial VO$_{2}$ films, highlighting the sensitivity of complementary metamaterial resonators to investigate the mesoscopic phase transition.

\section{\label{sec:level1}Complementary Vanadium Dioxide Metamaterial}
\subsection{\label{sec:level2}Terahertz transmission of pristine vanadium dioxide film}

The VO$_{2}$ film was deposited on a sapphire substrate by hybrid molecular beam epitaxy using a combinatorial growth \cite{brahlek2018frontiers, zhang2015wafer}. Details of the film growth process can be found in Appendix A. The frequency dependent transmission was measured using THz time domain spectroscopy (THz-TDS) \cite{Zhao:18}. THz pulses were generated using near infrared (800 nm) pulses generated by a Ti:sapphire laser with a pulse duration of 25 fs. The pulses illuminate a biased photoconductive antenna, and are detected using a similar photoconductive antenna gated by the 800 nm pulses.

THz-TDS transmission measurements of the pristine VO$_{2}$ film were performed prior to metamaterial integration. Fig. 1(a) shows the transmission of the VO$_{2}$ film at 315 K in the insulating monoclinic phase and at 360 K in the rutile metallic phase (Tc $\sim$340 K). Near unity transparency is observed close to room temperature for a 100 nm thick VO$_{2}$ film in the insulating state (we note that the transmission is referenced to a bare sapphire substrate). Referencing errors caused by birefringent mismatch between the sample and substrate reference along with artifacts arising from an etalon in the sapphire substrates results in small oscillations in the transmission above unity at some frequencies. In the metallic state, the transmission decreases to $\sim$58\%. This corresponds to a $\sim$42\% transmission amplitude modulation between the insulating and metallic states. We note that the flat spectral response in the metallic state arises from the short scattering time of the quasiparticles. The temperature dependent transmission amplitude at 0.47 THz is shown in the inset of Fig. 1. Hysteresis upon heating and cooling is observed as expected for a first order phase transition. Fitting the first derivative of the hysteresis loop to a Gaussian yields a transition temperature with heating of 339 K, and with cooling of $\sim$330 K, corresponding to a hysteresis width of 9 K. This dynamic change of transmission under different temperatures makes VO$_{2}$ a potential material for tunable THz bandpass filters\cite{huang2021switchable,li2021terahertz}.

\begin{figure}
\includegraphics[width=0.48\textwidth]{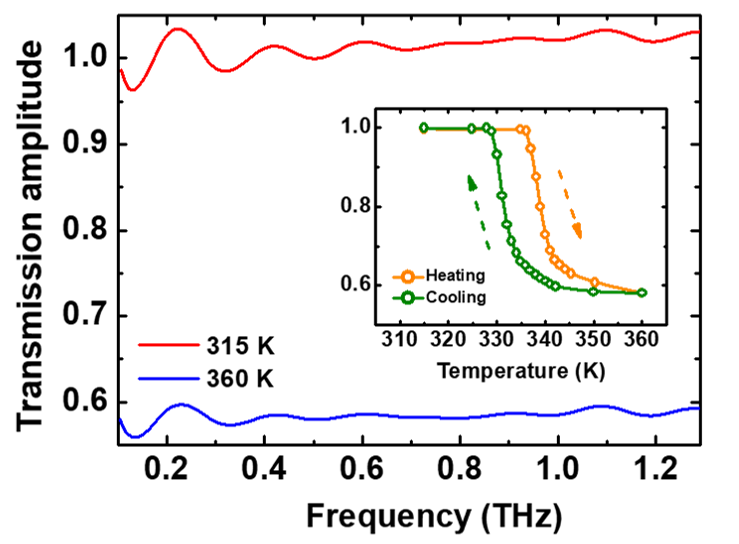}
 \caption{(a) Terahertz transmission amplitude versus frequency of the pristine VO$_{2}$ film in the monoclinic insulating state (red line, 315 K) and rutile metallic phase (blue line, 360 K). Inset is the THz transmission (at 0.47 THz) of the VO$_{2}$ film as a function of temperature across the IMT transition. The expected hysteresis is observed upon heating (orange dots – line is guide to eye) and cooling (green dots – line is guide to eye).}%
\end{figure}

\subsection{\label{sec:level2} Design, simulations, and experimental results}

Metallic SRR metamaterials exhibit a LC dipole resonance that can modify the transmission modulation of the VO$_{2}$ film. For conventional (i.e., not complementary -- see below)  split ring resonators (SRR) on top of a VO$_{2}$ film, there is a dip in transmission in the insulating state at the SRR resonance frequency (see Appendix A). Upon traversing the IMT, the  transmission will increase since the capacitive gaps of the SRRs are shorted by the metallic VO$_{2}$. In this case, the modulation depth is smaller in comparison to the pristine VO$_{2}$ film. However, based on Babinet’s principle \cite{zentgraf2007babinet,zhao2020terahertz}, a complementary SRR structure exhibits a peak in transmission due to the resonance in the insulating state of VO$_{2}$ which can be expected to decrease in the metallic state. Therefore, we investigate the transmission modulation of  complementary SRRs integrated with a VO$_{2}$ film for enhanced transmission modulation.

An array of complementary split ring resonators (CSRRs) was deposited on a VO$_{2}$ film using direct laser writing  and photolithography to construct a CSRR-VO$_{2}$ metamaterial (CSRR-VO$_{2}$ MM). Fig. 2(a) shows an optical image of the array. A 150 nm thick gold film (with a 10 nm Ti adhesion layer) was deposited on the VO$_{2}$ film. The lateral dimension of the CSRR is 45 $\mu$m with a periodicity of 50 $\mu$m, a linewidth of 5 $\mu$m, and slit widths of 3 $\mu$m on both branches. 

\begin{figure}
\includegraphics[width=0.48\textwidth]{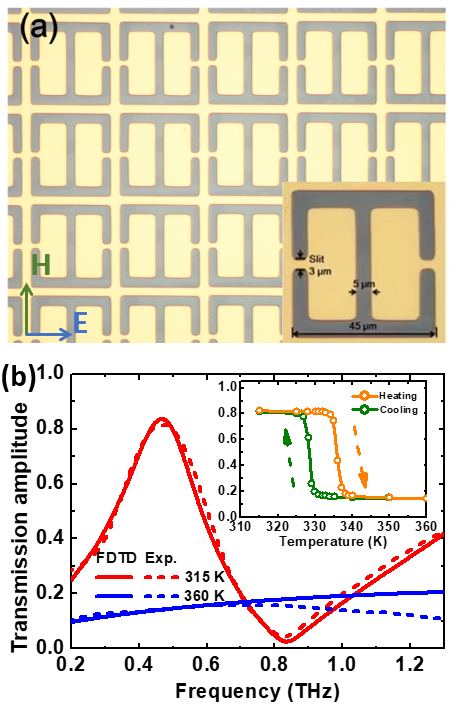}
 \caption{(a) Microscope image of the CSRR-VO$_{2}$ MM on VO$_{2}$ film where the inset shows the dimensions of the unit cell. (b) Experimental data (dashed lines) and simulations (solid lines) of the THz transmission of CSRR-VO$_{2}$ MM at 315 K (red) and 360 K (blue). Inset is the measurement of temperature-dependent transmission of CSRR-VO$_{2}$ MM at 0.47 THz. The solid lines in the inset are a guide to the eye.}%
\end{figure}

The frequency dependent transmission amplitude of the CSRR-VO$_{2}$ MM measured using THz-TDS is plotted as dashed lines in Fig. 2(b). The resonance peak occurs at 0.47 THz when the incident light is polarized parallel to the middle gap at 315 K, shown as the blue arrow in Figure 2(a). The transmission amplitude is 82.8\% (compared to the pristine VO$_{2}$ films, corresponding to an insertion loss of 17.2\% -- nonetheless the transmission is reasonably high). Conversely, upon fully traversing to the metallic state to 360 K, no resonant transmission  response is evident (blue line, Fig. 2(b)), and the transmission is 14.5\% at 0.47 THz. Thus, the modulation amplitude of the THz transmission at 0.47 THz upon crossing the IMT is 68.3\%, an increase of 62.4\% relative to the bare VO$_{2}$ (Fig. 1). 

Numerical simulations of the electromagnetic response of the CSRR-VO$_{2}$ MM were performed using CST Microwave Studio. The conductivity of gold was taken as 4.5×10$^7$ S/m. In the insulating state, the VO$_{2}$ was modeled using a Drude response with the complex permittivity given by \cite{cocker2010terahertz,berglund1969electronic} :
\begin {equation}
\varepsilon(\omega)=\varepsilon_\infty-\frac{\omega_p^2}{ \omega(\omega+i\gamma)}
\end {equation}
where the unscreened plasma frequency $\omega_p$ is given by $\sqrt{\emph{n}_{d}e^2/(\varepsilon\emph{m}*)}$, $\emph{n}_{d}$ is the carrier density, $\emph{e}$ is electron charge, $\emph{m}$* is the carrier effective mass, $\gamma$ is the scattering frequency ($\gamma = \emph{e}$/$\mu\emph{m}$*, where $\mu$ is the mobility), and $\varepsilon_{\infty}$=10 is the high frequency permittivity. The parameters used in the simulations are shown in Table 1 and the corresponding simulation results are shown as red solid lines in Fig. 2(b). For the metallic state of VO$_{2}$, an equivalent carrier density of 10$^{21}$ cm$^{-3}$ at 360 K was used as determined from the THz-TDS measurements of the VO$_{2}$ film shown in Fig. 1. The parameters for both the insulating and metallic states in Table 1 are consistent with those in published work \cite{liu2012terahertz}. Good agreement with experiment is obtained (Fig. 2(b)) in the metallic state, with a simulated transmission amplitude of 14.7\% at 0.47 THz. In addition, the inset in Fig. 2(b) shows the temperature dependence of transmission amplitude at 0.47 THz for the CSRR-VO$_{2}$ MM during both heating and cooling. A significant change in the transmission upon heating did not occur until 335 K. Indeed, a dramatic reduction of the transmission occurs within a 2 K span from 335 K to 337 K. Similarly, upon cooling, the transmission recovery occurs at a temperature 5 K lower than the turning point of heating. 

The mechanism for the increase in transmission modulation going through the IMT can be partially understood by examining the electric fields and current distributions at the resonant frequencies using full-wave electromagnetic simulations. Fig. 3 shows both the electric field and surface current at the resonant frequency (0.47 THz) when VO$_{2}$ is in the insulating state (315 K) and metallic state (360 K). As shown in Fig. 3(a), the electric field in the insulating state is concentrated in the central vertical gap while in the metallic state, this region is electrically shorted, reducing the resonant enhancement of the transmission. The surface currents on the CSRR-VO$_{2}$ MM for both states are shown with colored arrows in Fig. 3(c) and (d). The circulating current in Fig. 3(c) is indicative of the lowest energy LC mode of the structure. This is the current path that leads to the enhanced transmission on resonance. However, in the metallic state, the VO$_{2}$ film allows current to flow across the central gap, effectively shorting the capacitive region and diminishing the LC mode, as shown in Fig. 2(b), resulting in a low transmission amplitude. 

\begin{figure}
\includegraphics[width=0.48\textwidth]{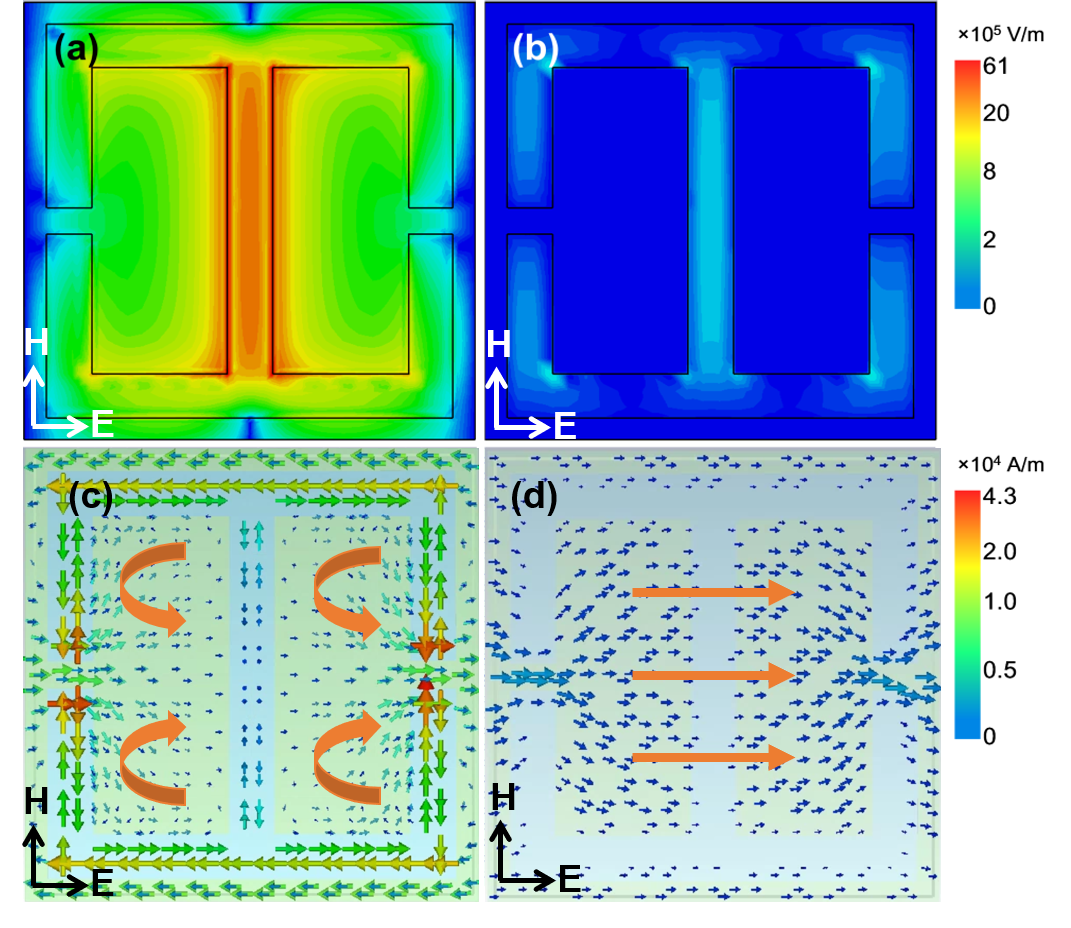}
 \caption{Simulated electric field of the CSRR-VO$_{2}$ MM in the (a) insulating state and (b) metallic state, and the corresponding simulated surface current distributions in the (c) insulating and (d) metallic state at 0.47 THz.}%
\end{figure}

\subsection{\label{sec:level2} Temperature dependence of transmission}

Having presented the functional response of the CSRR-VO$_{2}$ MM, we now consider the temperature dependence of the transmission upon crossing the IMT in greater detail. Fig. 4(a) and (b) show the experimental transmission spectra at different temperatures during heating and cooling, respectively. Notably, with increased temperature, the resonant frequency dramatically redshifts over a narrow temperature interval, decreasing from 0.47 THz at 334 K to 0.38 THz at 336 K, as shown in Fig. 4(a). A further increase of temperature results in a fully screened resonance, leading to a flat transmission spectrum at 338 K and 360 K. Similar blueshifts can also be observed in Fig. 4(b) during cooling after the resonance peak reappears from 328 K to 315 K. 

To clarify the origin of the resonant frequency shifts, we need to consider the frequency dependent dielectric response of the VO$_{2}$ film over the temperature range of the IMT. We first consider the Drude response in Eq. (1). Based on the published work, there is a large change of carrier concentration $n_{d}$ from $\sim$10$^{20}$ cm$^{-3}$ in the insulating state to $\sim$10$^{21}$ cm$^{-3}$ in the metallic state, while the mobility $\mu$ increases from 1 to 10 cm$^{2}$/(V$\cdot$s), and the effective mass $m$* decreases from 7 $m_{0}$ to $m_{0}$ \cite{park2013measurement}. Based on the parameters mentioned above, we are able to determine $\omega_p$ and $\gamma$ in both the insulating and the metal state according to Eq. (1).

\begin{table*}
\caption{\label{tab:table3}Permittivity parameters used for the VO$_{2}$ simulations.}
\begin{ruledtabular}
\begin{tabular}{cccccc}
% &\multicolumn{2}{c}{$D_{4h}^1$}&\multicolumn{2}{c}{$D_{4h}^5$}\\
 Process & Temperature (K) & Electron density (cm$^{-3}$) & Mobility (cm$^{2}$/V$\cdot$s) & Effective mass (m$_{0}$) & $\varepsilon_\infty$\\ \hline
         & $315$ & $1.0\times10\textsuperscript{20}$ & $1.0$ & $7$ & $10$ \\
         & $335$ & $1.2\times10\textsuperscript{20}$ & $1.2$ & $6$ & $100$ \\
 Heating & $336$ & $2.3\times10\textsuperscript{20}$ & $2.0$ & $5$ & $500$ \\
         & $337$ & $5.0\times10\textsuperscript{20}$ & $6.0$ & $3$ & $10$ \\
         & $338$ & $7.0\times10\textsuperscript{20}$ & $7.0$ & $2$ & $10$ \\
         & $360$ & $1.0\times10\textsuperscript{21}$ & $10.0$ & $1$ & $10$ \\
         & $330$ & $7.0\times10\textsuperscript{20}$ & $7.0$ & $2$ & $10$ \\
Cooling  & $329$ & $3.5\times10\textsuperscript{20}$ & $5.0$ & $3$ & $200$ \\
         & $328$ & $1.5\times10\textsuperscript{20}$ & $1.5$ & $5$ & $250$ \\
         & $327$ & $1.1\times10\textsuperscript{20}$ & $1.1$ & $6$ & $50$ \\
         & $315$ & $1.0\times10\textsuperscript{20}$ & $1.0$ & $7$ & $10$ \\
\end{tabular}
\end{ruledtabular}
\end{table*}

In order to better understand the redshift of the resonance during IMT, it is useful to express Eq. (1) in terms of the real and imaginary parts of the permittivity:

\begin {equation}
\varepsilon(\omega)=(\varepsilon_\infty-\frac{\omega_p^2}{ \omega^2+\gamma^2})+i\frac{\omega_p^2\gamma}{\omega(\omega^2+\gamma^2)}=\varepsilon_1+i\varepsilon_2
\end {equation}

We first consider a constant $\varepsilon_{\infty}$ during the IMT. When the temperature increases, $n_d$ increases while $m$* decreases, corresponding to an increase of $\omega_{p}$. Specifically, $\omega_{p}$ increase from $\sim$200 THz in the insulating state to $\sim$1800 THz in the metallic state. At the same time, $\mu$ increases from 1 to 10, which corresponds to $\gamma$ decreasing from 250 THz in the insulating state to 170 THz in the metallic state. The combined increase of $\omega_{p}$ and decrease of $\gamma$ yields a significant decrease of $\varepsilon_{1}$ at 0.47 THz from 10. We also simulated the response with $\varepsilon_{\infty}$ =10 while $n$, $m$* and $\mu$ are individually changed (see Appendix C). The simulation results show that the resonant frequency cannot be tuned over 0.1 THz range under constant $\varepsilon_{\infty}$ =10 even when the transmission amplitude decreases to 0.4. This clearly shows that the Drude response with constant $\varepsilon_{\infty}$ across the IMT cannot account for the experimentally observed frequency shift. Therefore, we must consider additional effects in the VO$_{2}$ film during the IMT. Specifically, the experimentally observed redshift indicates that $\varepsilon_{\infty}$ increases in the vicinity of the IMT. 

Fig. 4(c) shows the simulation results which includes a significant increase in $\varepsilon_{\infty}$ upon heating, using the parameters in Table 1. Simulations using these values capture the redshift of the resonance observed in experiment. The similar blueshift observed during the cooling process can also be explained by gradually decreasing $\varepsilon_{\infty}$, as shown in Fig. 4(d) and Table 1.

\begin{figure*}
\includegraphics[width=0.8\textwidth]{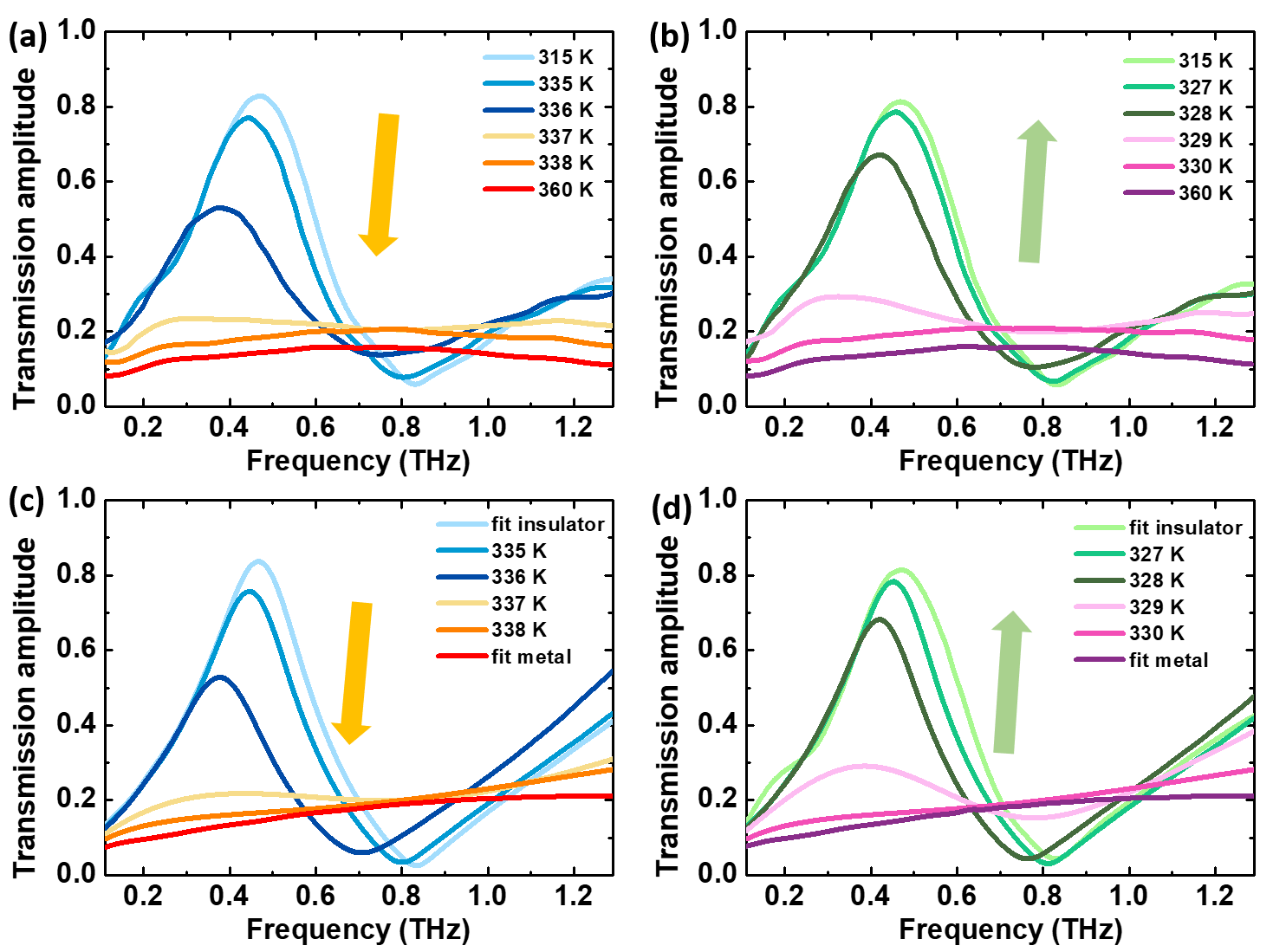}
 \caption{Experimental and numerical simulation spectra of the VO$_{2}$ complementary metamaterial upon crossing the IMT. (a) Experimental heating and (b) cooling results, with corresponding numerical simulations shown in (c) for heating and (d) for cooling using the values from Table 1.}%
\end{figure*}

Given the  excellent agreement between simulation and experiment presented in Fig. 4, it is important to understand why the permittivity of VO$_{2}$ would exhibit a significant increase during the IMT. In principle, structural changes could lead to huge changes in the permittivity, such as in ferroelectric materials, e.g. BaTiO$_{3}$ and CaCu$_{3}$Ti$_{4}$O$_{12}$ \cite{litvinchuk2003optical,dang2013flexible}. On the other hand, scanning near-field infrared microscopy images of polycrystalline VO$_{2}$ films indicate a percolative process during the IMT \cite{qazilbash2009infrared,qazilbash2009infrared,liu2013anisotropic,he2009high}. Metallic puddles emerge, sparsely embedded in the insulating matrix of VO$_{2}$, with the volume fraction of the metallic puddles increasing with temperature. At high enough temperatures, the metallic puddles coalesce into a uniform rutile metallic phase (unity metal volume fraction). This process is similar to increasing the concentration of conducting nanoparticles in an otherwise dielectric medium. This can be describe using the Maxwell-Garnett effective medium theory (MG-EMT) model 
 \cite{jepsen2006metal}. In this model, the effective permittivity of a dielectric matrix with a metallic volume fraction $f$ of conducting particles is expressed as:

\begin{equation}
    \varepsilon_{eff}=\varepsilon_{i}\frac{\varepsilon_{m}(1+2f)-\varepsilon_{i}(2f-2)}{\varepsilon_{i}(2+f)+\varepsilon_{m}(1-f)}
\end{equation}

\noindent where $\varepsilon_{i}$ is the permittivity in the insulating state, and $\varepsilon_{m}$ is the permittivity in the metallic state. Based on the simulated model, in both the insulating state and metallic state, we used $\varepsilon_{i}$=10 and $\varepsilon_{m}$=-90 between 0.1 to 1.3 THz, in agreement with the published work\cite{emond2017natural,qazilbash2007mott}. Fig. 5(a) shows the relationship between the metallic volume fraction $f$ and the effective permittivity $\varepsilon_{eff}$ based on Eq. (3). It is clear that there is a significant increase of the effective permittivity from 10 to $\sim$500 when the metallic volume fraction increases to $\sim$0.62 along with a sudden decrease and stabilization to -90 when $f$ increases to 1. Combined with the relationship between $f$ and temperature of the VO$_{2}$ film during IMT in the published work\cite{hilton2007enhanced}, we are able to estimate the effective permittivity of the VO$_{2}$ film at our experimental temperatures (colored stars in Fig. 5(b)). We also include the permittivities used in the simulations at 0.5 THz, shown as the red hollow circles in Fig. 5(b). Reasonable agreement is obtained between the calculated and the simulated permittivity, consistent with a significant permittivity increase during the IMT. We noted that the calculated permittivity at 338 K using Eq. (3) is much lower than the simulated permittivity. This can be attributed to the huge decrease of the calculated permittivity from $\sim$400 to $\sim$-400 when $f$ is between 0.62 to 0.67. Thus, either a slight change of $\varepsilon_{m}$ or the $f$ chosen for 338 K would result in a significant change of permittivity at that temperature. Nonetheless, the trend of the change of permittivity during the IMT used in the simulation is in reasonable agreement with the MG-EMT model. We also calculated the optical conductivities at $\sim$0.5 THz (see Appendix D) and the values are also consistent with published work \cite{qazilbash2007mott}.

The Maxwell-Garnett model indicates that the percolation threshold of our VO$_{2}$ film is $\sim$0.65. The increases in the metallic volume fraction results in an effective increase in the surface area and narrowing distance between the of the capacitive region of the CSRR. These combined effects contribute an abnormally large capacitance. A significant permittivity increase occurs in close proximity to the percolation threshold $f_{t}$. This is precisely the behavior across the insulator-to-metal transition that results in a giant effective permittivity. This, as our simulations reveal, can account for the observed redshift of the LC resonance. An even higher value of the permittivity of $\sim$2700 was achieved with a composite comprised of graphite nanoplates and polyimide, though such a composite lacks the tunability of VO$_{2}$ \cite{he2009high}.

Above the percolation threshold, the metallic puddles in the VO$_{2}$ film coalesce into a uniform metallic film. As such, $\varepsilon_\infty$ recovers to the original value of 10 and the second term in Eq. (2) dominates, such that the permittivity jumps to an overall negative value \cite{qazilbash2007mott}. Consequently the metallic VO$_{2}$ film shorts the gaps of the CSRR, as shown in Fig. 4(b) and (d).

Our results indicate that, in addition to the enhancement of the transmission modulation across the IMT, integration of the CSRR MM is highly sensitive to the local permittivity. Unlike the very small capacitive gap in a typical SRR, the central ``wire'' in the complementary structure acts as the capacitive gap, markedly increasing the area of the resonator which probes the local dielectric environment, thereby increasing the overall sensitivity. Furthermore, much of the existing research on VO$_{2}$ films focused on effects perpendicular to the plane and thus on the length scale of the film thickness, typically limited to tens to hundreds of nanometers \cite{qazilbash2007mott}.  These scales are too small to observe a significant influence of the percolative effects. In contrast, the relatively large scale of the capacitive region of the CSRR-VO$_{2}$ MM structure avoids this limitation. A previous report exhibited this behavior \cite{liu2012terahertz}, but the complementary structure makes the effect significantly more prominent. That is, the evolution of the CSRR-VO$_{2}$ MM resonant frequency across the IMT provides straightforward evidence of the percolative behavior in our VO$_{2}$ film.

\begin{figure}
\includegraphics[width=0.48\textwidth]{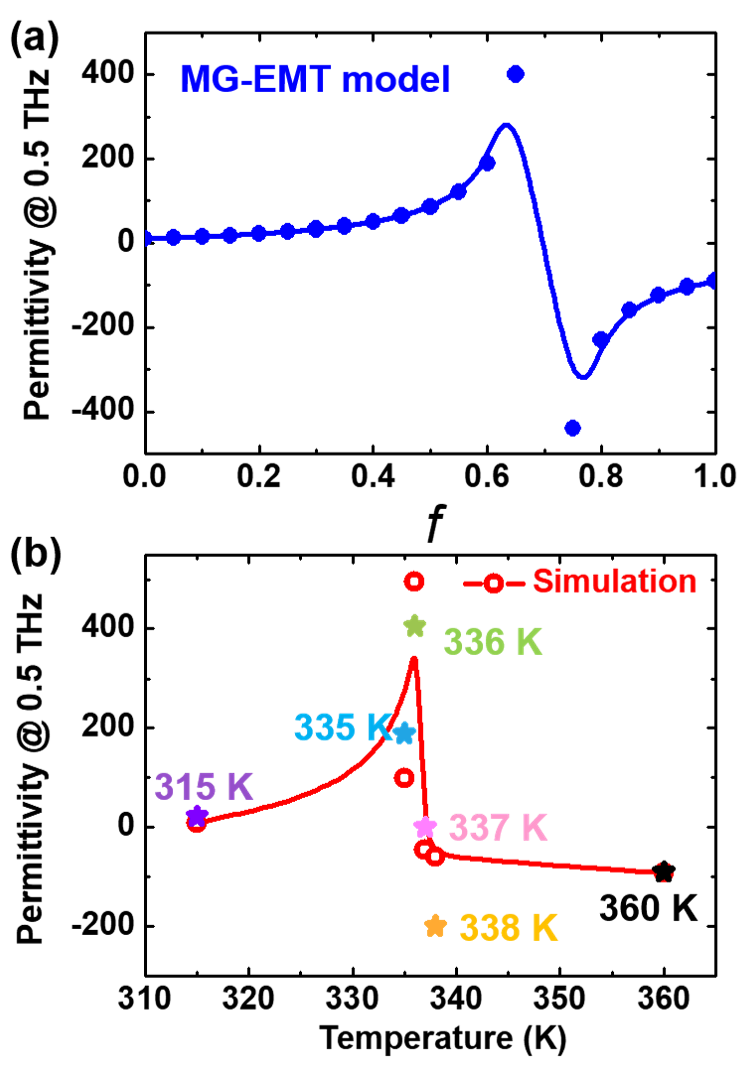}
 \caption{(a) The metallic volume fraction $f$ dependent permittivity of VO$_{2}$ based on Maxwell-Garnett effective medium theory model. (b) Comparison between the permittivity of VO$_{2}$ in the simulation (red hollow circles) and the calculation from (a) (colored stars).}%
\end{figure}

\section{\label{sec:level1}Conclusion}
We demonstrated metamaterial-enhanced dynamic material properties by integrating a complementary split ring resonator metamaterial directly on a VO$_{2}$ film. Relative to the IMT in a pristine VO$_{2}$ film, the modulation of the THz transmission (peaked at 0.47 THz) of our hybrid device is enhanced by 62.4\%. Moreover, the complementary SRR structure has a larger effective gap area than typical SRR geometries, providing a sensitive probe of the local electrodynamics of the VO$_{2}$. With increasing (decreasing) temperature, we observe a dramatic redshift (blueshift) in the resonant frequency of the CSRR-VO$_{2}$ MM structure in a narrow temperature range in the vicinity of the transition temperature. The significant increase of the VO$_{2}$ permittivity during the onset of IMT explains the redshift of the resonance. Maxwell-Garnett effective medium theory provides reasonable agreement with experiment in describing the temperature dependence of the transmission during IMT. In summary, metamaterials and related metasurface constructs provide a powerful route to obtain dynamic enhancement and dielectric sensitivity that not only facilitates the development of tunable devices, but also provides a simple and effective means to interrogate the local electrodynamic properties of the materials with which they are integrated.

\begin{acknowledgments}
The work at Boston university is supported by National Science Foundation under Grant No. ECCS- 1810252. The work at UCSD is supported by ARO MURI Grant No. W911NF-16-1-0361. Film growth (HTZ and REH) was supported by National Science Foundation under Grant No. DMR-1352502 and the Penn State MRSEC program DMR-1420620. 
\end{acknowledgments}

\appendix

\section{VO$_{2}$ film growth}
The VO$_{2}$ film was deposited on a sapphire substrate by hybrid molecular beam epitaxy using a combinatorial growth approach. Here, control over the vanadium valence state was achieved by co-suppling elemental vanadium (valence state 0) and the vanadium containing alkoxide precursor vanadium-oxo-triisopropoxide (VTIP), in which vanadium assumes the valence state 5+. Optimal flux ratios were determined using a flux gradient method to establish a valence state library across the wafer during a calibration run. Relative tuning of the two fluxes allowed straight-forward balancing of the vanadium valence state to stabilize the vanadium 4+ state. achieving an epitaxial VO$_{2}$ film with excellent uniformity and record-high resistance change across the metal-to-insulator transition exceeding four orders of magnitude \cite{zhang2015wafer}.

Growth was carried out in DCA M600 MBE growth chamber and VO$_{2}$ thin films were grown on a 3-inch r-cut sapphire wafer co-supply of VTIP (vacuum distilled, trace metal impurity 4N, MULTIVALENT Laboratory) using a heated gas inlet system retrofitted to an effusion cell port of the MBE growth chamber and vanadium metal (4N, Ames Laboratory) evaporated from a high temperature Knudsen cell. The respective fluxes were adjusted and maintained via a proportional-integral-derivative (PID)-controlled adjustable leak valve and capacitance manometer on the VTIP gas inlet system and PID controlled Knudsen cell temperature. Sapphire wafer were cleaned in ultrasonic bath of acetone and then isopropanol prior to loading them into the MBE system and baked at 423 K for 2 h in the load lock chamber, subsequently transferred into the MBE growth chamber and heated to the film growth temperature of 623 K. Before deposition substrates were exposed for 20 min to an oxygen plasma (power 250 W, 9×10$^{-7}$ torr oxygen background pressure). The VO$_{2}$ film were grown at a rate of about 5Å min-1 in the presence of an oxygen plasma by co-supplying a vanadium atom flux of 4×10$^{12}$ cm$^{-2}$s$^{-1}$, calibrated by a quartz crystal monitor prior to growth, and matching the required VTIP flux determined from the experimental valence state library obtained from the valence state library. After growth the VO$_{2}$ films were cooled in the presence of oxygen plasma.

\section{SRR \& CSRR structure}

\begin{figure}[hbt!]
\includegraphics[width=0.48\textwidth]{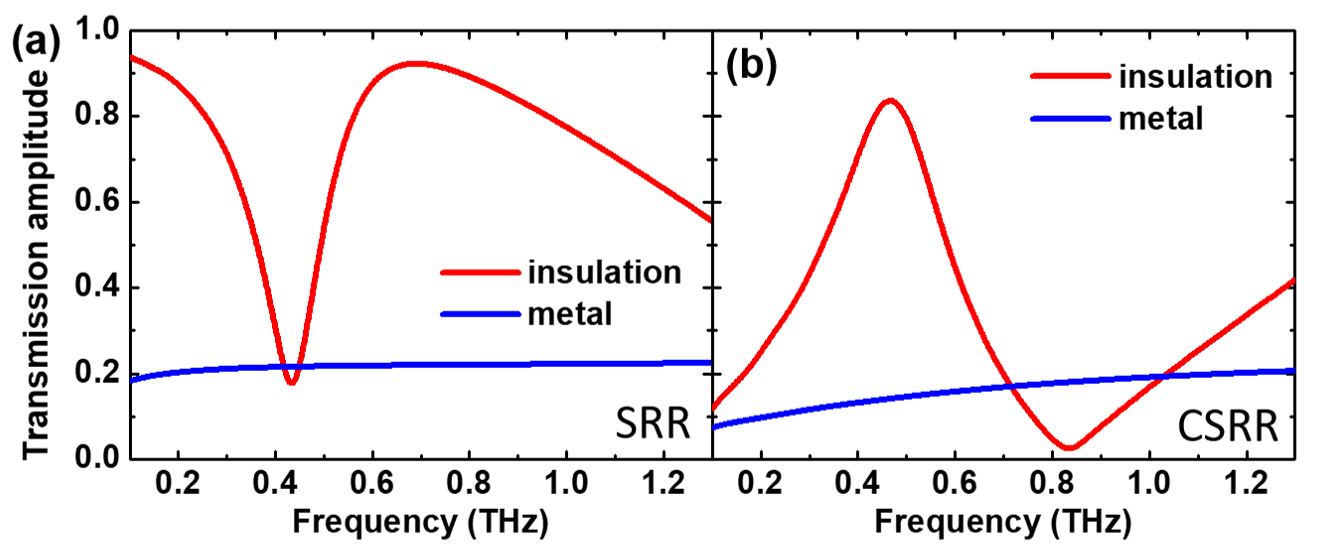}
 \caption{Simulations of transmission spectra of (a) SRR structure and (b) CSRR structure on VO$_{2}$ film in insulating (315 K) and metallic states (360 K).}%
\end{figure}

The LC resonance in a SRR structure manifests as a dip in transmission at 0.47 THz, decreasing the transmission to $\sim$0.17 when VO$_{2}$ is in insulating state, as shown in Fig. 6(a). In the metallic state, the significant increase in conductivity of the VO$_{2}$ film screens the LC resonance in SRR structure, resulting in a featureless transmission spectrum with amplitude of $\sim$0.2. Hence, the modulation amplitude across the IMT in VO$_{2}$ is small when using a SRR structure. However, based on Babinet’s principle, we are able to achieve transmission peak with amplitude of $\sim$0.83 at almost the same frequency using the CSRR structure, as shown in Fig. 6(b). In the insulating state for the CSRR, the transmission is $\sim$0.83 decreasing to $\sim$0.14 in the metallic state. Therefore, the modulation amplitude is higher using the CSRR structure in comparison to the SRR structure.

\section{Dependence of the resonant frequency on effective mass ($m$*), mobility ($\mu$) and carrier density ($n$)}

\begin{figure*}[hbt!]
\includegraphics[width=1\textwidth]{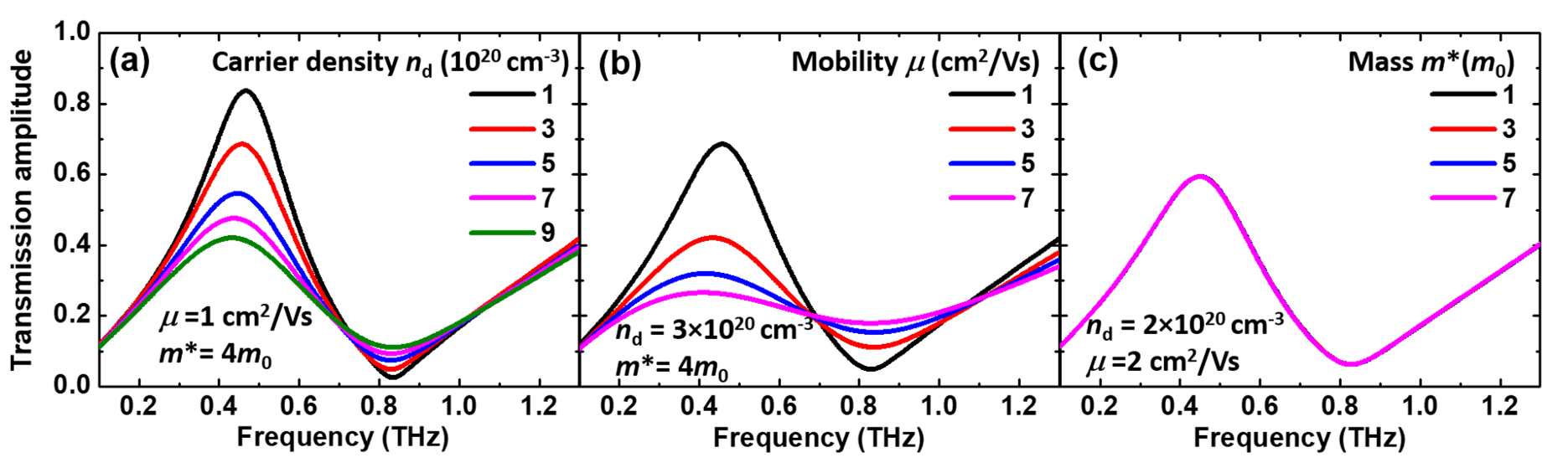}
 \caption{Dependence of transmission of CSRR-VO$_{2}$ MM on (a) effective mass ($m$*), (b) mobility ($\mu$) and (c) carrier density ($n_{d}$).}%
\end{figure*}

Based on the Drude model, the dielectric response of VO$_{2}$ is determined from the plasma frequency ($\omega_{p}$) and scattering frequency ($\gamma$), which depend on the  effective mass ($m$*), mobility ($\mu$) and carrier density ($n$) as described in the manuscript. During the IMT transition, $m$* decreases from 7 $m_{0}$ to $m_{0}$, $\mu$ increases from 1 to 10 while $n_{d}$ increases from $10^{20}$ to $10^{21}$ cm$^{-3}$. In Fig. 7,  we show simulations of the the transmission assuming a constant $\varepsilon_\infty$=10. Fig. 7 shows the dependence of the resonant frequency on $m$* (Fig. 7(a)), $\mu$ (Fig. 7(b)), and $n_{d}$ (Fig. 7(c)) while holding the other values constant. All three figures show that the resonant frequency exhibits a weak dependence on the three variables, in contrast to the experimental results. Thus, as described in the text and Fig. 4, an increase in $\varepsilon_{\infty}$ during the IMT determines the redshift of the resonant frequency.

\section{Conductivity calculated from the imaginary part of permittivity in simulation}

\begin{figure}[hbt!]
\includegraphics[width=0.45\textwidth]{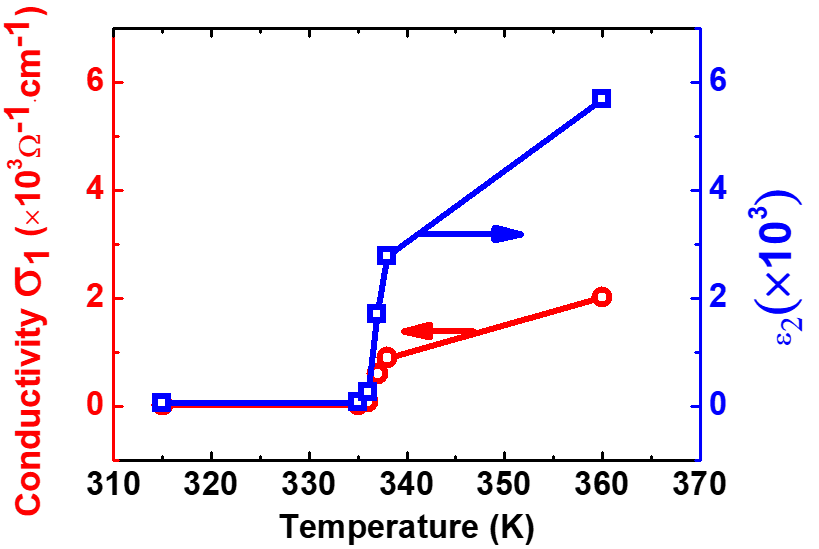}
 \caption{Imaginary part of permittivity ($\varepsilon_{2}$) extracted from simulations and corresponding real part of conductivity ($\sigma_{1}$) at 0.5 THz.}%
\end{figure}

Based on the parameters in Table 1, we extract the imaginary part of the permittivity at 0.5 THz from the simulation and calculated the optical conductivity \cite{qazilbash2007mott}:

\begin{equation}
    \sigma_{1}(\omega)=\frac{\omega\varepsilon_{2}(\omega)}{4\pi}
\end{equation}

The results are shown in Fig. 8. The optical conductivity for VO$_{2}$ in the metallic state is $\sim$2000 $\Omega^{-1}$cm$^{-1}$, which is similar to the conductivity in the published work \cite{qazilbash2007mott}. The conductivity is closely correlated with polarized metallic particles inside the material.


\clearpage

\nocite{*}

\bibliography{apssamp}% Produces the bibliography via BibTeX.

\end{document}